\documentclass[12pt,preprint]{aastex}
\usepackage{amsmath,amssymb}
\usepackage{hyperref}
\let\captionbox\relax
\usepackage{graphicx,caption,subcaption}
\usepackage{natbib}
\usepackage{multirow}
\newcommand{\mbf}[1]{\mathbf{#1}}
\newcommand{\fpar}[2]{\frac{\partial #1}{\partial #2}}

\begin{document}
\title{Magnetohydrodynamic modeling of solar coronal dynamics with an initial non-force-free magnetic field}

\author{A. Prasad, R. Bhattacharyya and Sanjay Kumar} 
\affil{Udaipur Solar Observatory, Physical Research Laboratory, Dewali, Bari Road, Udaipur-313001, India}

\begin{abstract}
The magnetic fields in the solar corona are generally neither force-free nor axisymmetric and have complex dynamics that are difficult to characterize. Here we simulate the topological evolution of solar coronal magnetic field lines (MFLs) using a magnetohydrodynamic model.
The simulation is initialized with a non-axisymmetric non-force-free magnetic field
that best correlates with the observed vector magnetograms of solar active regions (ARs). 
To focus on these ideas, simulations are performed for the flaring  AR 11283 noted for its complexity and well-documented dynamics.
The simulated dynamics develops as the initial Lorentz force pushes the plasma and facilitates
successive magnetic reconnections at the two X-type null lines present in the initial field. 
Importantly, the 
simulation allows for the spontaneous development of mass flow, unique among contemporary works, that preferentially reconnects field lines 
at one of the X-type null lines. Consequently, a flux rope consisting of low-lying twisted MFLs, which approximately traces the major polarity inversion line, undergoes an asymmetric monotonic rise. The rise is attributed 
to a reduction in the magnetic tension force at the region overlying the rope, resulting from the reconnection. A monotonic rise of the rope 
is in conformity with the standard scenario of flares.   Importantly, the simulated dynamics 
leads to bifurcations of the flux rope, which, being akin to the observed filament bifurcation 
in AR 11283, establishes the appropriateness of the initial field in describing ARs.  

\end{abstract}

\keywords{magnetohydrodynamics (MHD) -- Sun: activity -- Sun: corona -- Sun: flares -- Sun: magnetic fields -- Sun: photosphere}

\section{\label{sec:level1}Introduction}
The solar corona is generally treated as a magnetized plasma with high electrical conductivity whose evolution is determined by the equations of magnetohydrodynamics (MHD) {\citep{2014masu.book.....P}}. The large  
magnetic Reynolds number $R_M (vL/\eta$, in usual notations) of the corona allows the 
plasma to achieve its non-diffusive limit {\citep{2004psci.book.....A}}. Consequently, Alfv\'{e}n's  theorem of flux freezing is satisfied, ensuring magnetic field lines (MFLs) to remain tied to  fluid parcels during  evolution {\citep{1942Natur.150..405A}}. 
In contrast, instances of various eruptive events (flares, coronal mass ejections) 
occurring in the  corona are believed to be signatures of magnetic reconnection (MR): the topological 
rearrangement of MFLs along with the conversion of magnetic energy into heat and the kinetic energy of mass 
motion {\citep{2011LRSP....8....6S}}. To trigger MRs, the value of  $R_M$ needs to be lowered, which can be attained by a reduction in $L$, the length scale determining the gradient of the magnetic field. 
A vanishing $L$ can either pre-exist in a magnetic topology in the form of various two- and three-dimensional nulls and quasi-separatrix layers (QSLs),
or develop autonomously. The autonomous development 
is assured by 
the Parker's magnetostatic theorem \citep{1972ApJ...174..499P,1988ApJ...330..474P,1994ISAA....1.....P}, which claims 
that discontinuities of the magnetic field are ubiquitous in an infinite electrically conducting plasma in equilibrium, because the flux-freezing condition  and the equilibrium generally cannot be satisfied simultaneously by a magnetic field that is continuous everywhere. In an 
evolving fluid, MHD simulations establish the autonomous reduction of $L$ and the occurrence of 
consequent spontaneous reconnections as the fluid relaxes toward
a quasi-steady state under a nearly precise maintenance of flux freezing 
{\citep{2015PhPl...22a2902K,2016PhPl...23d4501K,2016ApJ...830...80K}}.   
However, in these studies, the initial bipolar magnetic fields were constructed analytically, reducing the natural complexity often observed in active regions (ARs).    

Without a direct measurement {\citep{2012LRSP....9....5W}}, the coronal 
fields are numerically extrapolated from the photospheric magnetic fields based on a  force-free assumption; i.e. exert no Lorentz force is exerted {\citep{2013SoPh..288..481R}}.  
The corresponding analytical expressions of the force-free field (in spherical polar coordinates) are often obtained under the assumption of axisymmetry  \citep{1990ApJ...352..343L,2014ApJ...786...81P}. 
However, such a symmetry generally does not exist in the real case, as many vector magnetograms depict the photospheric magnetic fields to be morphologically complex. 

In force-free coronal field extrapolations, the use of nonlinear-force-free-field (NLFFF) models where the field line twist is a function of position but constant along a field line \citep{2008JGRA..113.3S02W} is customary. The model is well known to give good fits to magnetograms, yielding a realistic coronal field line topology. For instance, \citet{2013ApJ...779..157G} performed NLFFF extrapolation for NOAA 10720 and found coronal 
flux ropes wrapped around by a particular set of QSLs, whereas \citet{2012ApJ...747...65I} found
that in AR 10930, the X-ray intensity associated with the sigmoid is more sensitive to the strength of the electric current rather than the twist of the field
lines. Furthermore, many recent works \citep{2013ApJ...771L..30J, 2013ApJ...779..129K, 2014Natur.514..465A, 2014ApJ...788..182I, 
2015ApJ...803...73I, 2015ApJ...810...96S, 2016ApJ...817...43S} have realistically modeled the coronal eruptive dynamics by using MHD simulations initialized with NLFFFs (see also a review by \citet{2016PEPS....3...19I}).

Although simulations based on NLFFF are capable of reproducing realistic dynamics, 
 further issues need to be addressed. Specifically, the force-free approximation is based on the assumption of negligible plasma $\beta$ in the corona, where $\beta$
is the ratio of gas pressure to magnetic pressure. However, according to \citet{2001SoPh..203...71G}, it is only in the mid-corona---sandwiched
between the photosphere and the upper corona---that the plasma $\beta$ is negligible.
To mitigate this problem, a `preprocessing' technique is often performed on the data to remove the force in the vector magnetogram and obtain a suitable boundary condition for NLFFF extrapolations \citep{2006SoPh..233..215W,2014SoPh..289...63J}.

An alternative to NLFFF extrapolation is an approach
based on a magnetic field model of the corona derived from the variational principle of the minimum energy dissipation rate \citep{2008SoPh..247...87H,2008ApJ...679..848H,2010JASTP..72..219H}.
It is shown that the extrapolated field lines from the resulting non-force-free-field (NFFF) model reasonably agree with the observed coronal loops and MFLs extrapolated using NLFFF. Central to the NFFF extrapolations is the double-curl Beltrami field obtained by  \citet {2007SoPh..240...63B}, which is a dissipating self-organized state. \citet{2011PhPl...18h4506K} have also shown the same double-curl Beltrami field to be the minimum total energy state of a two-fluid magnetoplasma. One particular analytical solution of the double-curl equation can be formulated by superposing two linear-force-free fields (LFFFs) characterized by constant twists everywhere cf. \citet{1998PhPl....5.2609S} \& \citet{1998PhRvL..81.4863M} for details.  \citet{2011PhPl...18h4506K} utilized such formalism to obtain NFFFs with twisted MFLs resembling coronal loops. A NFFF relevant to observations of AR 11283 was recently obtained by  \citet{2016PhPl...23k4504P} by using the field in conjunction with 
semi-analytical constructs where non-axisymmetric solutions are best correlated with vector magnetograms.  Interestingly, the obtained NFFF yields a magnetic morphology where regions of the sharpest magnetic field 
gradient were found to be cospatial with the flaring site. 
To further establish NFFFs in describing the corona, it is 
imperative to assess the credibility of the obtained magnetic field in reproducing the observed evolution of ARs. Specifically, in this paper, we perform numerical simulations with the non-axisymmetric NFFF as the initial condition and  compare the evolution with the recorded  dynamics of AR 11283. Importantly, unlike contemporary works that require a prescribed flow to induce dynamics   \citep{2003ApJ...585.1073A,2010ApJ...708..314A}, here the inherent non-zero Lorentz force serves that purpose.

To incorporate possible  spontaneous reconnections {\citep{2016ApJ...830...80K}}, 
the simulations are performed in congruence with the the magnetostatic theorem. For this purpose, an  
incompressible thermally homogeneous magnetofluid with infinite electrical conductivity is allowed  to relax to a quasi-steady state via viscous relaxation \citep{2010PhPl...17k2901B,2014PhPl...21e2904K,2015PhPl...22h2903K}
satisfying

\begin{eqnarray}
\label{stokes}
& & \frac{\partial{\bf{v}}}{\partial t} 
+ \left({\bf{v}}\cdot\nabla \right) {\bf{ v}} =-\nabla p
+\left(\nabla\times{\bf{B}}\right) \times{\bf{B}}+\frac{\tau_a}{\tau_\nu}\nabla^2{\bf{v}},\\  
\label{incompress1}
& & \nabla\cdot{\bf{v}}=0, \\
\label{induction}
& & \frac{\partial{\bf{B}}}{\partial t}=\nabla\times({\bf{v}}\times{\bf{B}}), \\
\label{solenoid}
& &\nabla\cdot{\bf{B}}=0, 
\end{eqnarray}  
in usual notations. The equations (\ref{stokes})-(\ref{solenoid}) are 
written in dimensionless form, with the following normalization

\begin{eqnarray}
\label{norm}
& &{\bf{B}}\longrightarrow \frac{{\bf{B}}}{B_0},\\
& &{\bf{v}}\longrightarrow \frac{\bf{v}}{v_a},\\
& & L \longrightarrow \frac{L}{L_0},\\
& & t \longrightarrow \frac{t}{\tau_a},\\
& & p  \longrightarrow \frac{p}{\rho {v_a}^2}. 
\end{eqnarray}

The constants $B_0$ and $L_0$ are generally arbitrary,
but can be fixed by the magnetic field strength and size of the
system. Also, $v_a \equiv B_0/\sqrt{4\pi\rho_0}$ is the Alfv\'{e}n
speed and $\rho_0$ is the constant mass density. The constants $\tau_a$ and
$\tau_\nu$ have dimensions of time, and represent the Alfv\'{e}n transit
time ($\tau_a=L_0/v_a$) and viscous diffusion time scale ($\tau_\nu=
L_0^2/\nu$), respectively, with $\nu$ being the kinematic viscosity.
The ratio  $\tau_a/\tau_\nu$ can be interpreted as an effective viscosity, which,
along with other forces, determines the dynamics.
The pressure perturbation, $p$, about a thermodynamically uniform ambient state 
satisfies the elliptic boundary value problem, which is
generated by imposing the discretized incompressibility constraint (equation \eqref{incompress1}) 
on the discrete integral form of  the momentum equation (equation \eqref{stokes}); cf.\citep{2010PhPl...17k2901B} and the references therein. 
An identical procedure involving the gradient of an auxiliary potential in the induction equation (equation \eqref{induction}) is employed 
to keep ${\mathbf{B}}$ solenoidal,  see \citet{2010ApJ...715L.133G} and \citet{2013JCoPh.236..608S} for details.

If released from non-equilibrium, the initial Lorentz force will 
push the plasma and generate mass motion which is dissipated by the viscosity. 
With the flux freezing condition satisfied, the magnetic field cannot decay out completely but 
the terminal kinetic energy is expected to be zero because of viscous dissipation. 
In the corresponding  numerical simulation scales inevitably 
become under-resolved  and numerical artifacts
such as spurious oscillations are generated through the employed numerical
techniques. These underresolved scales can be removed by utilizing an appropriate
numerical diffusivity of the 
non-oscillatory finite difference algorithm. In the
literature, such calculations  relying on the non-oscillatory numerical
diffusivity are referred to as the Implicit Large Eddy Simulations (ILESs),
\citep{grinstein2007implicit}. The computations performed in this work
are in the spirit of ILESs and the simulated MRs, being intermittent 
in space and time, mimic physical reconnections in high-$R_M$ fluids.

The rest of the paper is organized as follows. Section 2 highlights the important steps in constructing the 
non-axisymmetric non-force-free magnetic field.   
In Section 3, the numerical model is briefly discussed. The 
results of the simulation are presented in Section 4. Finally, in Section 5 we summarize the results and highlight key findings.

\section{{\bf{Initial magnetic field}}}
The  initial NFFF $\mathbf{B}$  is generated by 
superposing two linear force-free fields, ${\mathbf{B}}^1$ and  ${\mathbf{B}}^2$ \citep{2007SoPh..240...63B,2011PhPl...18h4506K}:

\begin{equation}
\mathbf{B} = \mathbf{B}^1  + \kappa \mathbf{B}^2. \quad 
\label{e:nff}
\end{equation}
The constant $\kappa$  is the relative amplitude of the superposing fields.
Being linear-force-free, the  $\mathbf{B}^1$ and $\mathbf{B}^2$ satisfy 
\citep{1956PNAS...42....1C,1957ApJ...126..457C}
\begin{equation}
\nabla\times\mbf{B}^i = \alpha_i \mbf{B}^i, 
\label{e:ff1}
\end{equation}
where $i=1,2$. The constant $\alpha_i$ represents the torsion coefficient \citep{2012PPCF...54l4028P} for the field ${\mathbf{B}}^i$
and $\alpha_2=(1+\gamma) \alpha_1$ with $\gamma$ measuring the deviation
 of $\mathbf{B}$ from being linear-force-free. 
A convenient expression of  $\mbf{B}^i$
in spherical polar coordinates is
\begin{equation}
\mbf{B}^i=\frac{1}{\alpha_i} \nabla\times\nabla\times\psi^i \mbf{r}+\nabla\times\psi^i \mbf{r},
\label{e:ck}
\end{equation}
 where ${\bf{r}}$ is the radial vector.  The scalar eigenfunction $\psi^i$ satisfies the Helmholtz equation $(\nabla^2+\alpha_i^2)\psi^i=0$ and can be expressed as 
\begin{equation}
\psi^i_{lm}(r,\theta,\phi)=[C_{lm}^1 j_l(\alpha_i r)+C_{lm}^2 y_l(\alpha_i r)]P_l^m(\cos\theta)\exp(im\phi),
\label{e:psi}
\end{equation}
where $j_l(r,\theta,\phi)$ and $y_l(r, \theta,\phi)$ are the spherical Bessel functions of the first and second kinds respectively and $P_l^m$ is an associated Legendre polynomial of degree $l$ and order $m$;
$C_{lm}^1$ and $C_{lm}^2$ are constants.
The individual components of $\mbf{B}^i$ are 
\begin{subequations}
\begin{align}
B^i_r(r,\theta,\phi)&=\frac{-1}{\alpha_i r}\left[\frac{1}{\sin\theta}\fpar{}{\theta}\left(\sin \theta \fpar{\psi^i}{\theta}\right)+\frac{1}{\sin^2\theta}\fpar{^2\psi^i}{\phi^2}\right], \label{br}\\
B^i_\theta(r,\theta,\phi)&=\frac{1}{\alpha_i r}\fpar{}{r}\left(r\fpar{\psi^i}{\theta}\right)+\frac{1}{\sin\theta}\fpar{\psi^i}{\phi},\label{bt}\\
B^i_\phi(r,\theta,\phi)&=\frac{1}{\alpha_i r \sin \theta}\fpar{}{r}\left(r\fpar{\psi^i}{\phi}\right)-\fpar{\psi^i}{\theta}\label{bp}.
\end{align}
\label{bf}
\end{subequations}
Further, to make $B^i$ physically meaningful without having infinite energies at $r=\infty$ \citep{1978SoPh...58..215S}, we bound it above by confining the MFLs in a spherical shell of radius $r_0 \leq r \leq r_1$
 \citep{2016PhPl...23k4504P}. Consequently,
\begin{align}
\psi^i_{lm}(r,\theta,\phi)=&[j_l(\alpha_i r)y_l(\alpha_i r_1)-j_l(\alpha_i r_1)y_l(\alpha_i r)] P_{lm}(\cos\theta)\exp(im\phi),
\end{align}
where $r_0$ is the inner radius 
of the shell selected to avoid singularity at the origin.  

To simulate the solar AR field using $\mathbf{B}$, we follow the mathematical steps described in \citet{2014ApJ...786...81P}, which provide fast and reasonably good fits to vector magnetograms. The procedure involves generating of two-dimensional vector magnetogram templates from  three-dimensional analytical solutions 
computed over the volume spanned by the spherical shell. To achieve this, a plane 
tangential to the shell at a location $r^\prime$ is constructed, and the components of $\mathbf{B}$ are computed on this plane. The plane 
is identified by a specific AR cutout whose orientation can further be changed by varying  the two Euler angles $\theta'$ and $\psi'$; see Figure \ref{f:geometry}. Finally, the best correlated $\mathbf{B}$ at the cutout is obtained by maximizing  

\begin{equation}
  c=\frac{\langle(\mathbf{B}\cdot\mathbf{B}_O)| \mathbf{B}_O|\rangle}{\langle|\mathbf{B}|^3
\rangle^{1/3}\langle|\mathbf{B}_O|^3\rangle^{2/3}},
\label{fitting}
\end{equation}
which correlates $\mathbf{B}$ with the field $\mathbf{B}_O$ obtained from the magnetogram.
 Explicitly, the parameters $\alpha_1$, $l$, $m$, $\kappa$ and $\gamma$, and the variables $r_0$, $r_1$, $r^\prime$, $\theta'$, $\psi'$ 
are obtained by maximizing $c$. Using this technique, \citet{2014ApJ...786...81P} analyzed vector magnetograms for AR 10930 spanning a period of three days with two X-class flares. Their estimates for the changes in free energy and relative helicity of the AR before and after the flares were found to be in good agreement with those obtained from other NLFFF extrapolations.
In this work, we use vector magnetograms from the Heliospheric Magnetic Imager (HMI)\citep{2012SoPh..275..229S} on 
board the Solar Dynamics Observatory (SDO)\citep{2012SoPh..275....3P}.
 The HMI (hmi.sharp\_cea\_720s data series) provides full-disk vector magnetograms of the Sun with a spatial resolution of $0''.5$ and a temporal cadence of 12 minutes. The magnetogram  $\mathbf{B}_O$ is initially remapped to a Lambert cylindrical equal-area projection and then transformed into heliographic coordinates \citep{1990SoPh..126...21G}. The selected vector magnetogram is for the AR 11283 on 2011 September 6 at 21:00 (Figure \ref{f:mag} a), the same AR as discussed in \citet{2016PhPl...23k4504P} but observed at an earlier time. The AR produced an X- class flare on 2011 September 6, peaked around 22:20 hr, and was followed by a coronal mass ejection \citep{2013ApJ...765...37F,2016A&A...591A.141J}. The initial magnetic field is constructed by using the magnetogram data at 21:00 hr, which documents that the magnetic field is morphologically complex with a strong bipolar region and other weaker diffuse field regions (Figure \ref{f:mag}). 

The details of the parameters for the best correlated non-axisymmetric NFFF are given in Table \ref{t:init} and the corresponding MFLs 
(top and side views) are shown in Figure \ref{f:mag} b and Figure \ref{f:int}.
\begin{table}[ht]
\centering
\resizebox{0.75\textwidth}{!}{%
\begin{tabular}{|c|c|c|c|c|c|c|c|c|c|c|}
\hline
 $c$ & $\alpha_1,\alpha_2$ & $l$ & $m$ &$\kappa$&$\gamma$ & $r_0$&$r^\prime$ & $r_1$ & $\theta'$ & $\psi'$ \\ \hline
0.54    & 9.67, 13.23    & 3   & 2   &0.125& 0.375 & 0.1& 0.34 & 1     & 1.83 & 0.13          \\ \hline
\end{tabular}%
}
\caption{Parameters used in construction of the initial non-axisymmetric non-force-free field. Here, the  parameters $\alpha_1$, $\alpha_2$, $l$, $m$, $\kappa$ and $\gamma$ determine a particular mode while $r_0$, $r^\prime$ $r_1$, $\theta'$ and $\psi'$ determine the extent of the computation domain.}
\label{t:init}
\end{table}
The figures are overlaid with the contours of $\mathbf{B}_z$ on the $z=0$ plane, which confirm the overall morphological similarity of the constructed field with the magnetogram in terms of having strong bipolar and weaker diffuse field regions.
The MFLs connecting the strong polarity regions are twisted loops, having the appearance of a forward S, and hence  resembles  sigmoid emission in AR 11283 seen 
by the Atmospheric Imaging Assembly (AIA) 94 \AA~ channel \citep{2013ApJ...771L..30J}.
The field ${\mathbf{B}}$ has two-dimensional X-type nulls (Figure {\ref{f:int2}}) constituting the two null lines, marked by X1 and X2. The location of the nulls is determined using the technique reported in {\citet{2015PhPl...22h2903K}}.
To further analyze the initial field line geometry, in Figure \ref{f:initial_ropes}(a), we identify four sets of twisted MFLs (colored red, yellow, green and gray), 
structurally resembling a magnetic flux rope, which is a stack of magnetic flux surfaces (MFSs) spanned by twisted field lines \citep{schrijver2009heliophysics}. To closely examine the resemblance, Figure \ref{f:initial_ropes}(b) 
plots a group of MFLs (in color red) with circularly distributed initial footpoints. The circle is 
marked by the curve P with its center located at $(x, y, z) = (0.23, 0.66, 0)$.
Notably, the figure confirms the MFLs to be to be twisted by approximately a half turn (angle $\pi$) about the central axis.
 Such visual confirmation of twist in field lines has also been done in \citet{2014ApJ...786L..16J}.
 After tracing the loop trajectories in the computational domain, the locus of the endpoints of the MFLs generate a second closed 
curve P1 on the $z=0$ plane. Since field lines cannot intersect each other, the group of MFLs generating the two closed curves must have a single common axis, which is identified as the central axis of the rope. Further, MFLs join the closed curves  and are tangential to a surface, making it a MFS \citep{2010PhPl...17k2901B}. By changing 
the radius of the circle, a co-axial stack of MFSs is generated (Figure \ref{f:initial_ropes}(c)) which, within the accuracy of the field line integration scheme, is identified as a magnetic flux rope. Further analysis (not reported) finds a threshold for the radius of the circularly distributed initial points, above which the loci of the endpoints are no longer closed curves and do not give rise to a flux rope.

\section{Numerical model}
To solve the  MHD equations (\ref{stokes})-(\ref{solenoid}), we utilize the 
well established magnetohydrodynamic numerical model EULAG-MHD {\citep{2013JCoPh.236..608S}}, 
which is an extension of the hydrodynamic model EULAG predominantly used in atmospheric and climate research
\citep{Prusa20081193}. For the completeness, here we
mention  only important features of the EULAG-MHD and  
refer the readers to  \citet{2013JCoPh.236..608S} and references therein for 
detailed discussions. The model is based on the spatio-temporally second-order accurate 
non-oscillatory forward-in-time multidimensional positive definite advection 
transport algorithm, MPDATA {\citep{2006IJNMF..50.1123S}}. Importantly, the MPDATA has a proven dissipative property, which is intermittent and adaptive, allowing the formation of underresolved scales in field variables for a selected grid resolution. 
In the magnetofluid undergoing viscous relaxation, sharpening of the magnetic field gradient is 
expected to be unbounded, ultimately leading to MRs at locations where the separation of 
non-parallel field lines approaches the grid resolution. The MR process
per se is underresolved, but effectively regularized by the local
second-order residual dissipation of MPDATA, which is sufficient to sustain the monotonic nature of the solution. Being intermittent and adaptive,
the residual dissipation facilitates the model performing ILESs that mimic  the action of explicit subgrid-scale turbulence models {\citep{2006JTurb...7...15M}} whenever the concerned advective field is underresolved. Such
ILESs performed with the model have already been successfully utilized
to simulate magnetic reconnections (MRs) to understand their role in the development of various magnetic structures in the solar corona {\citep{2016ApJ...830...80K,2015PhPl...22a2902K}}.
The simulations presented  continue to rely on the effectiveness of ILES in regularizing the onset of MRs.

\section{Simulation results}
\begin{figure}[h!]
  \centering
    \includegraphics[width=0.5\linewidth]{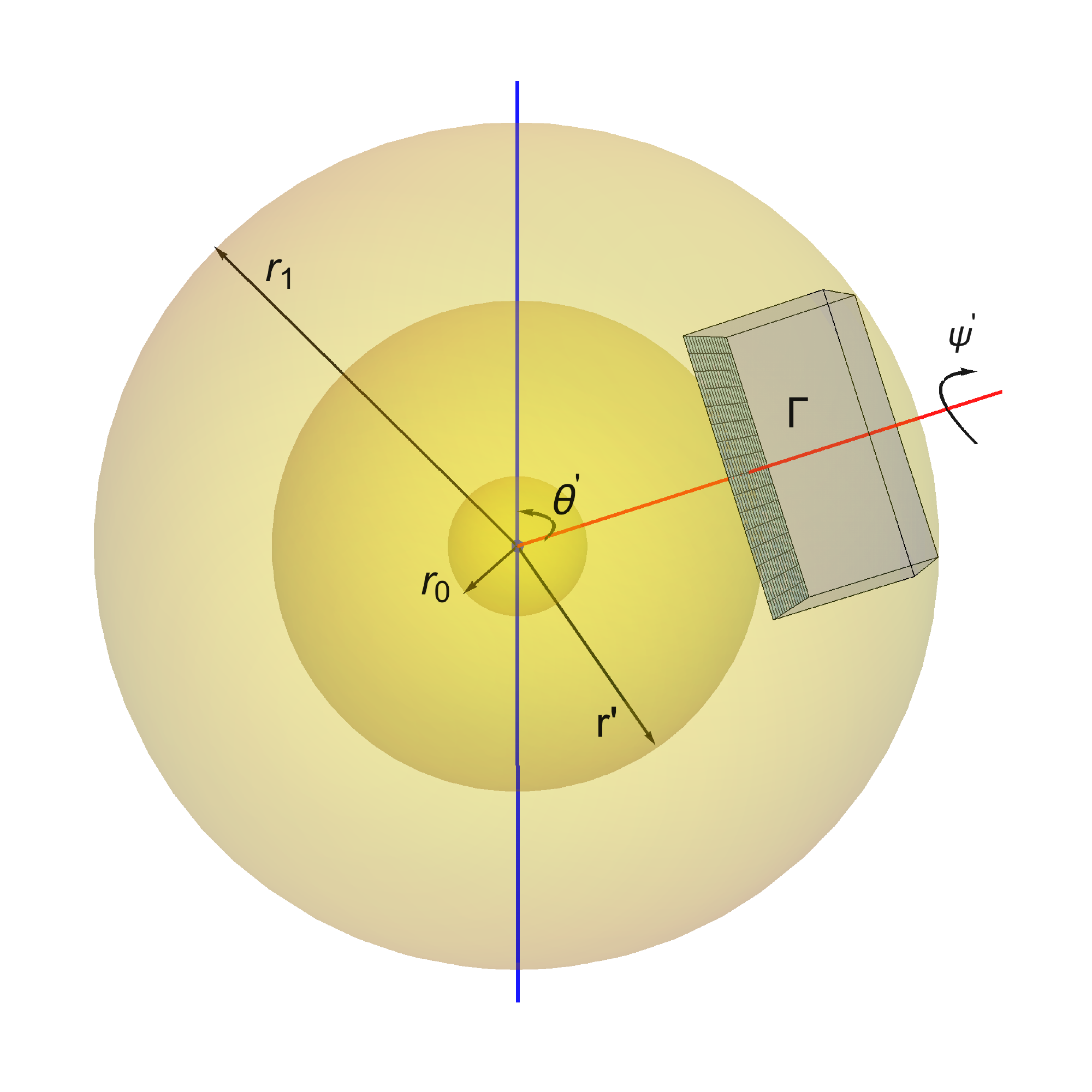}
  \caption{ Schematic of the geometrical construct utilized to best correlate the non-force-free magnetic field with magnetograms. The magnetogram cutout is
represented by the  bottom surface of the parallelepiped $\Gamma$, which  is the computational domain. In addition, the orientation of the surface
is  varied by changing the two Euler angles $\theta'$ and $\psi'$ to obtain the best correlation.}
  \label{f:geometry}
\end{figure}

\begin{figure}[h]
  \centering
  \begin{subfigure}[]{0.45\textwidth}
    \centering
    \includegraphics[width=1\linewidth]{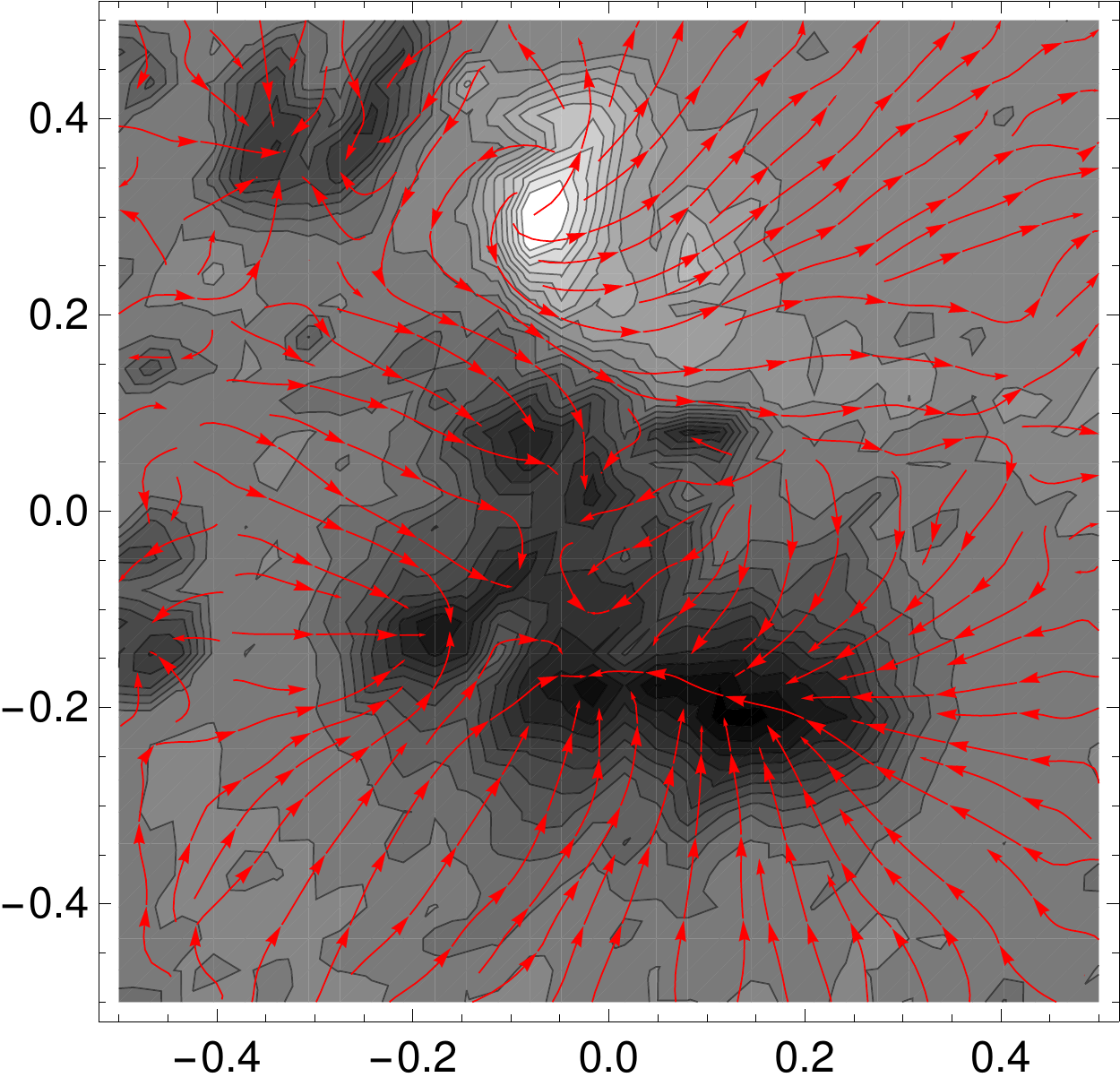}
    \caption{}
    \label{flow1a}
  \end{subfigure}
\quad
  \begin{subfigure}[]{0.45\textwidth}
    \centering
    \includegraphics[width=1\linewidth]{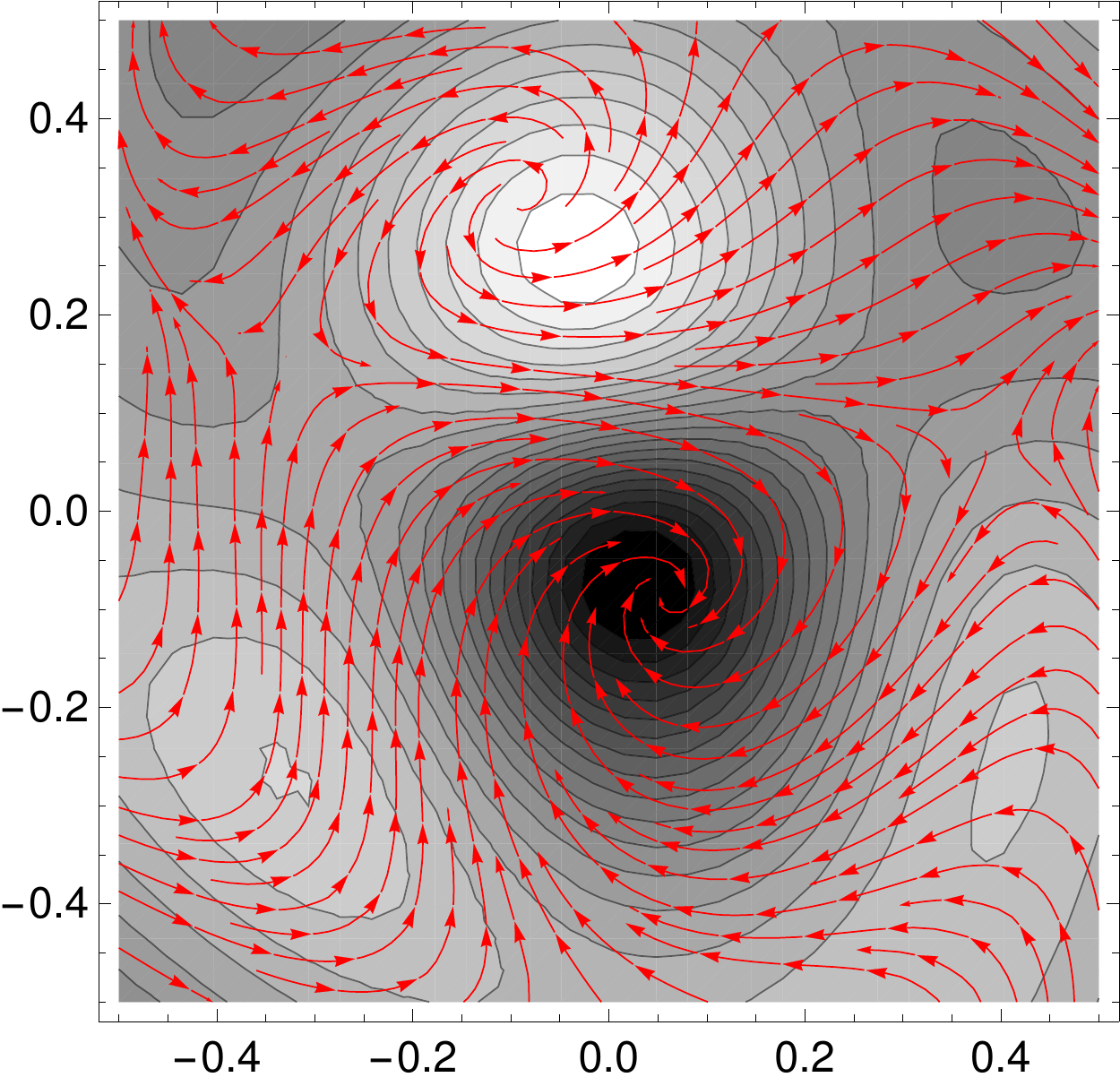}
    \caption{}
    \label{flow1b}
  \end{subfigure}
  \caption{(a) Selected region of the photospheric vector magnetogram of AR 11283 on 2011, September 6, 21:00 hr obtained from SDO/HMI. (b) Corresponding best-fit magnetogram using the non-force-free field. The contours of $B_z$ represent the strength of the vertical magnetic field whereas the streamlines depict the horizontal components.}
  \label{f:mag}
\end{figure}

\begin{figure}[h]
    \centering
    \includegraphics[width=\textwidth]{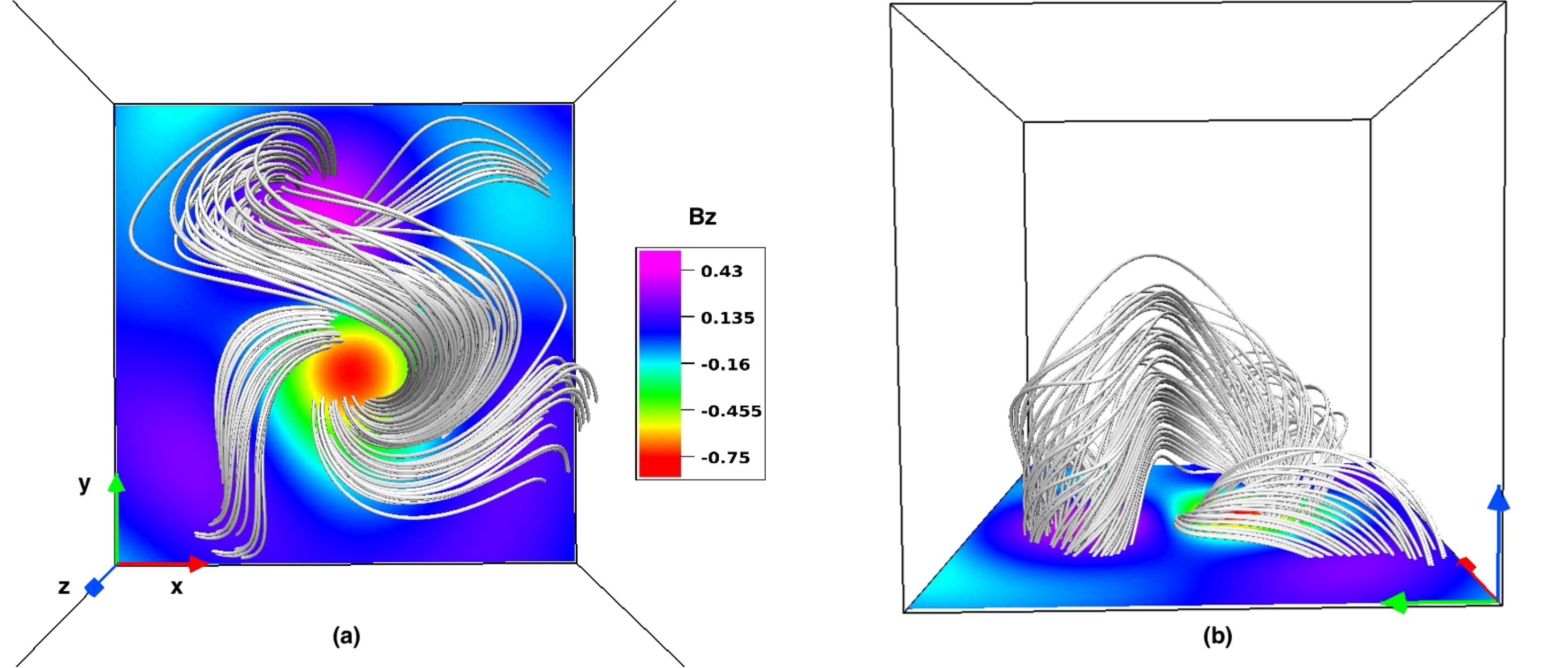}
    \caption{Initial field line topology.
    Panels (a) and (b) show the top and side views, respectively. The plots are overlaid with contours of $B_z$ at the $z=0$ plane. The presence of two strong opposite polarity regions (located in the central part) along with weak polarity regions (situated at the corners), leading to a complex field line topology, is notable. Also, the twisted nature of the MFLs is evident.}
  \label{f:int}
\end{figure}

\begin{figure}[h]
    \centering
    \includegraphics[width=\textwidth]{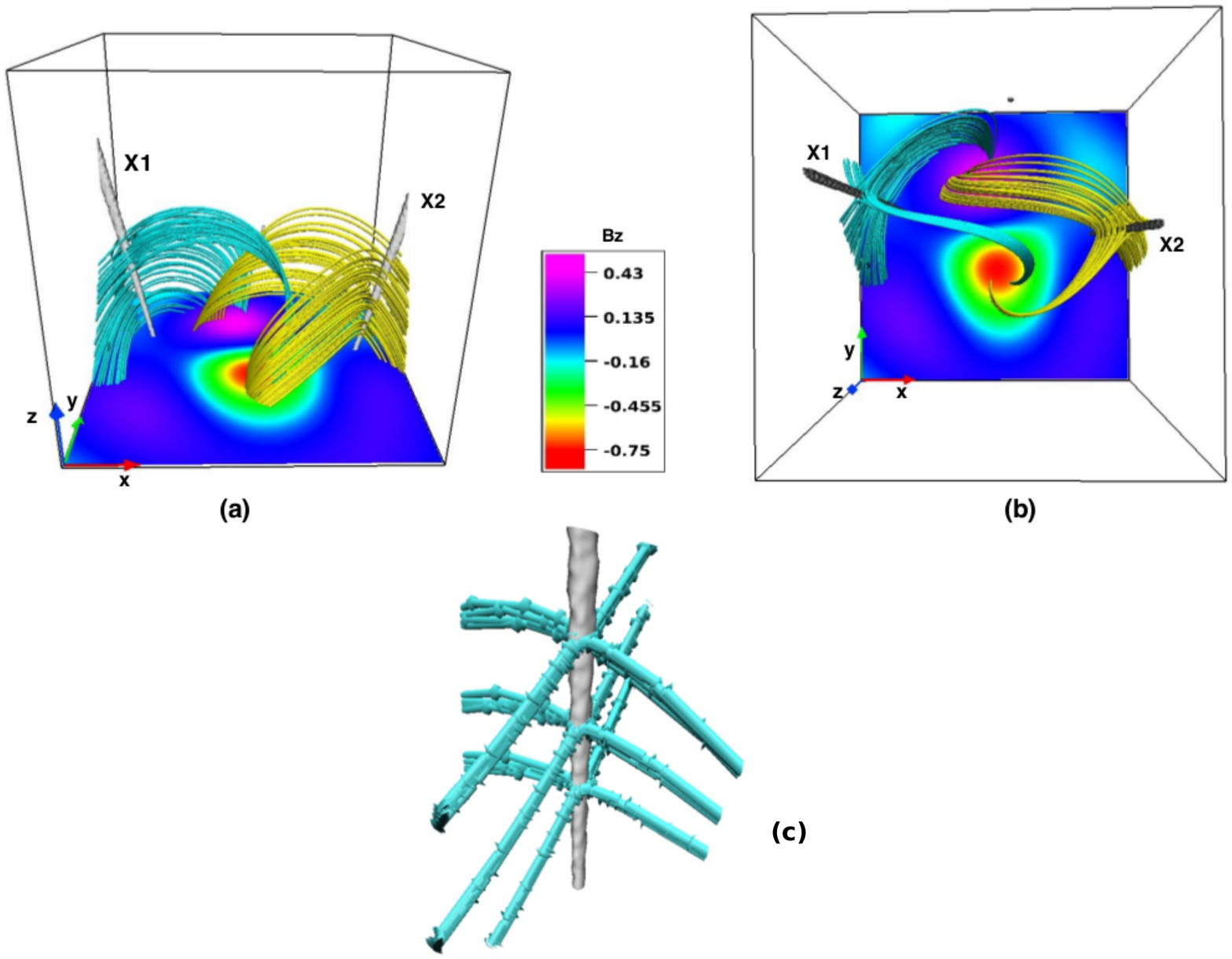}
   \caption{Side (panel (a)) and top (panel (b)) views of two-dimensional X-type magnetic nulls (in color gray) present in the initial field. A sequence of such nulls makes up the two 
null lines (marked by X1 and X2). For clarity, in panel (c) we plot MFLs near a section of the X2 line. The X-type geometry of the MFLs is evident.  }
 \label{f:int2}
\end{figure}

\begin{figure}[h]
  \centering
  \begin{subfigure}[]{0.47\textwidth}
    \centering
    \includegraphics[width=1\linewidth]{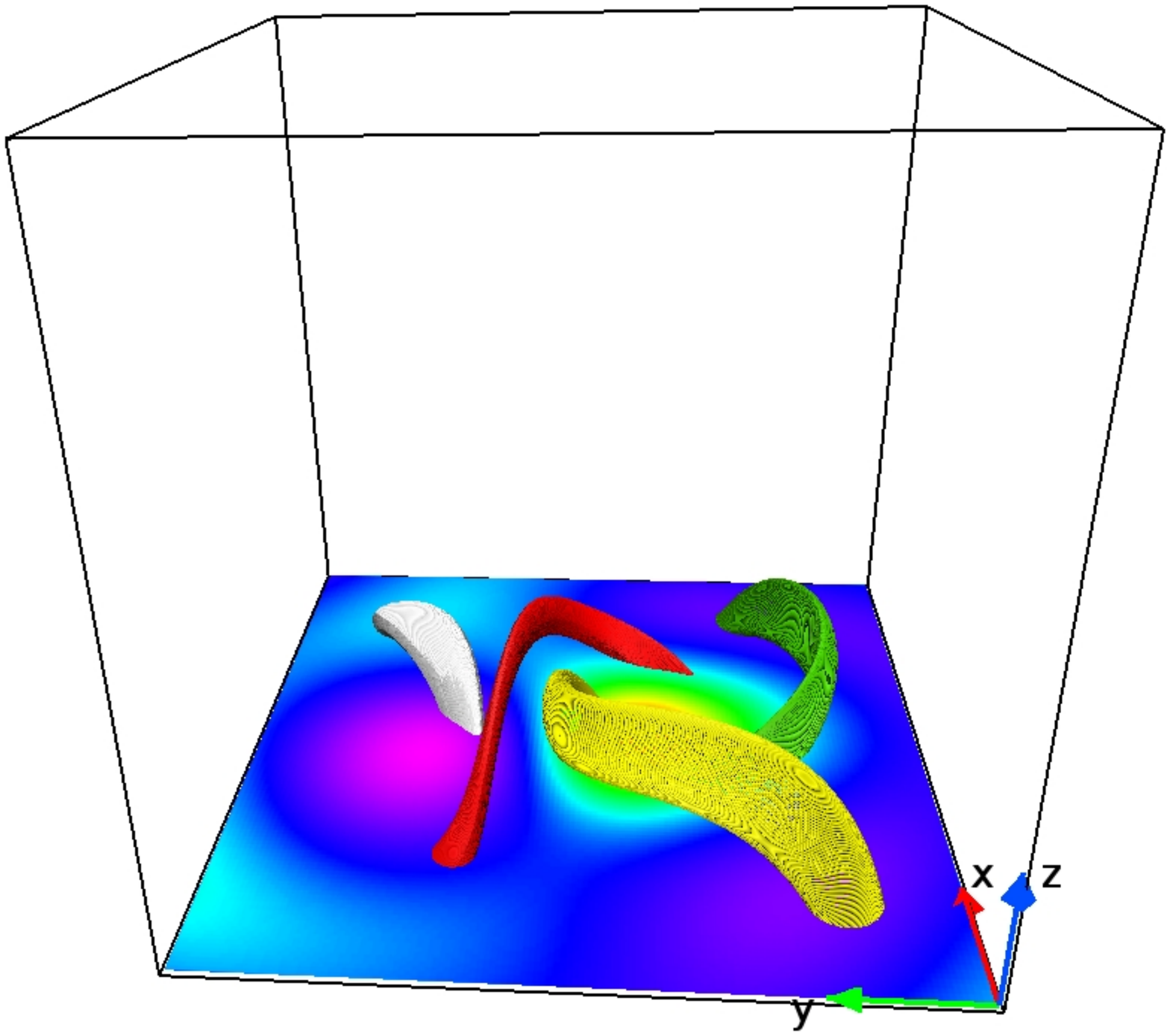}
    \caption{}
    \label{mrp}
  \end{subfigure}
\quad
  \begin{subfigure}[]{0.47\textwidth}
    \centering
    \includegraphics[width=1\linewidth]{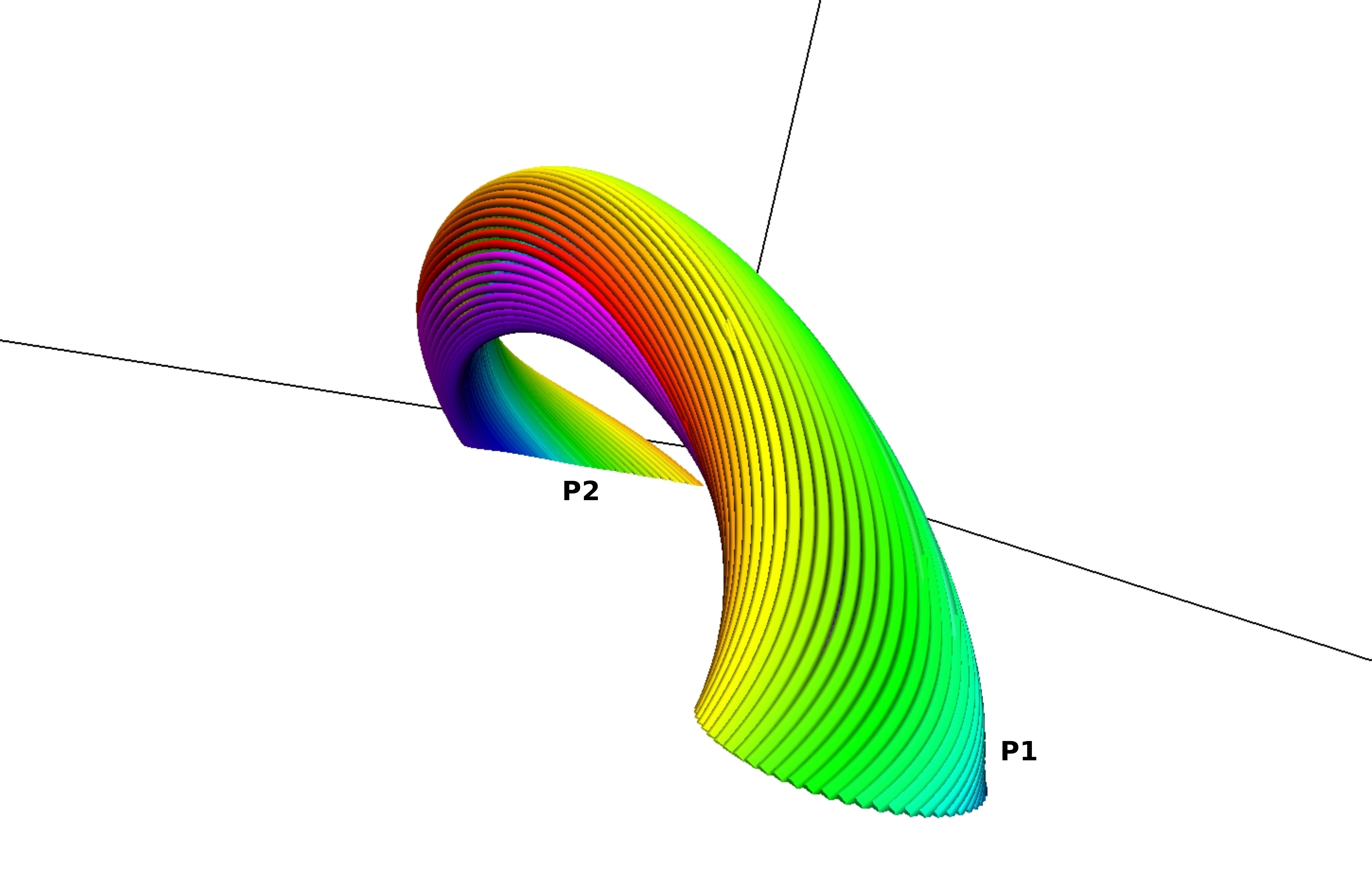}
    \caption{}
    \label{irp}
  \end{subfigure}
  \quad
  \begin{subfigure}[]{0.47\textwidth}
    \centering
    \includegraphics[width=1\linewidth]{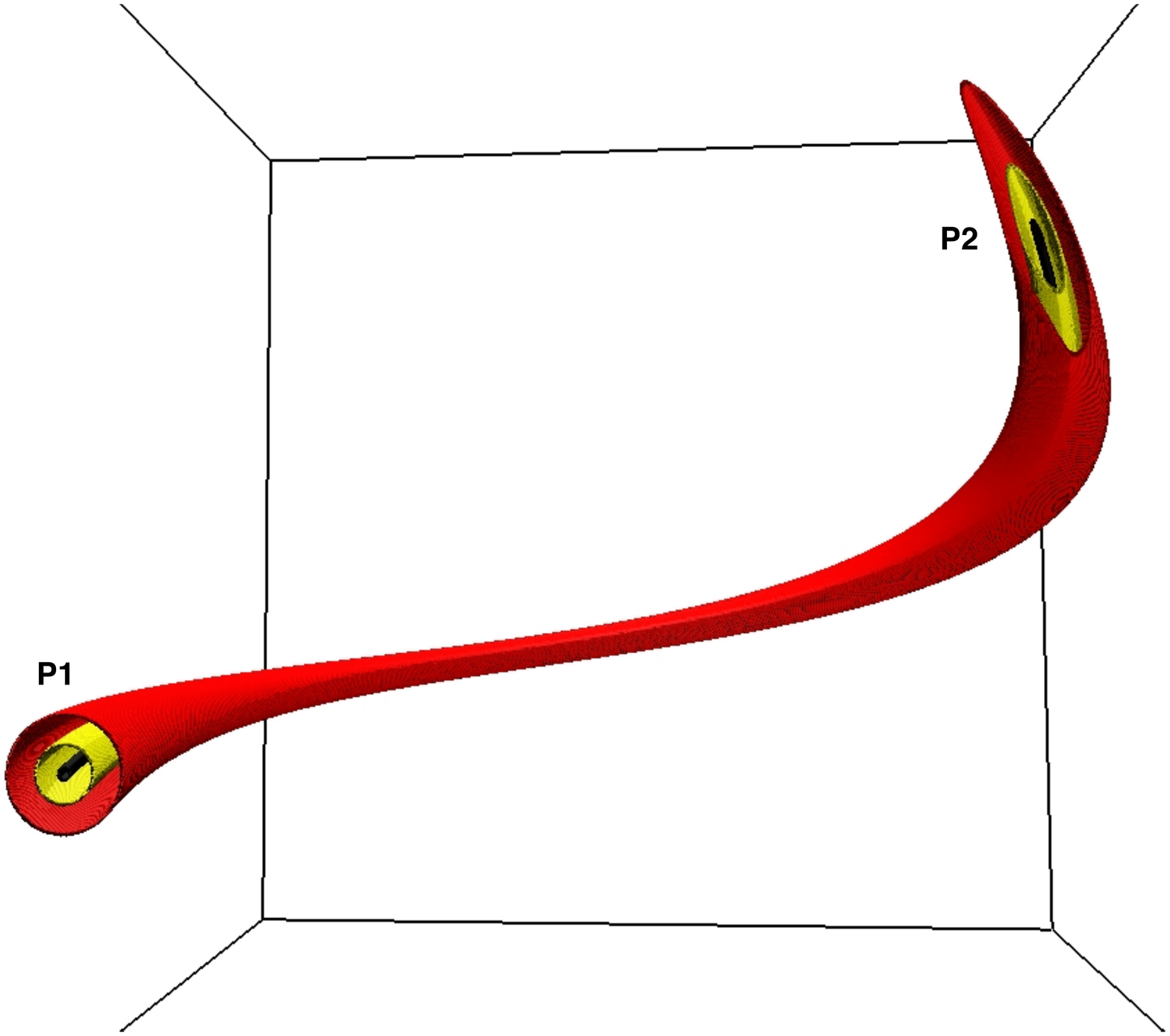}
    \caption{}
    \label{rp}
  \end{subfigure}
  \caption{ Panel (a) illustrates four flux ropes  (in red, yellow, green and gray) in the initial field ${\mathbf{B}}$. {Panel (b), presents a closeup of the red colored rope 
by plotting the constituent field lines. Further, field lines originating from different locations are represented by separate colors. Notably, a given colored field line rotates
considerably before intersecting the $z=0$ plane. The rotation visually confirms the twisted nature of the MFLs constituting the rope.} 
To further validate the rope, in panel (c) we display three co-axial local magnetic flux surfaces.  The surfaces are made of MFLs plotted with their initial points distributed on  circles  having three different 
radii and a common center situated at $(x, y) = (0.23, 0.66)$ on the $z= 0$ plane. The corresponding endpoints of the MFLs also form closed curves, enabling 
the group of MFLs with a given initial radius to constitute a flux surface. A stack of such flux surfaces
generates a rope.}
  \label{f:initial_ropes}
\end{figure}

\begin{figure}[h]
  \centering
  \begin{subfigure}[]{0.6\textwidth}
    \centering
    \includegraphics[width=\textwidth]{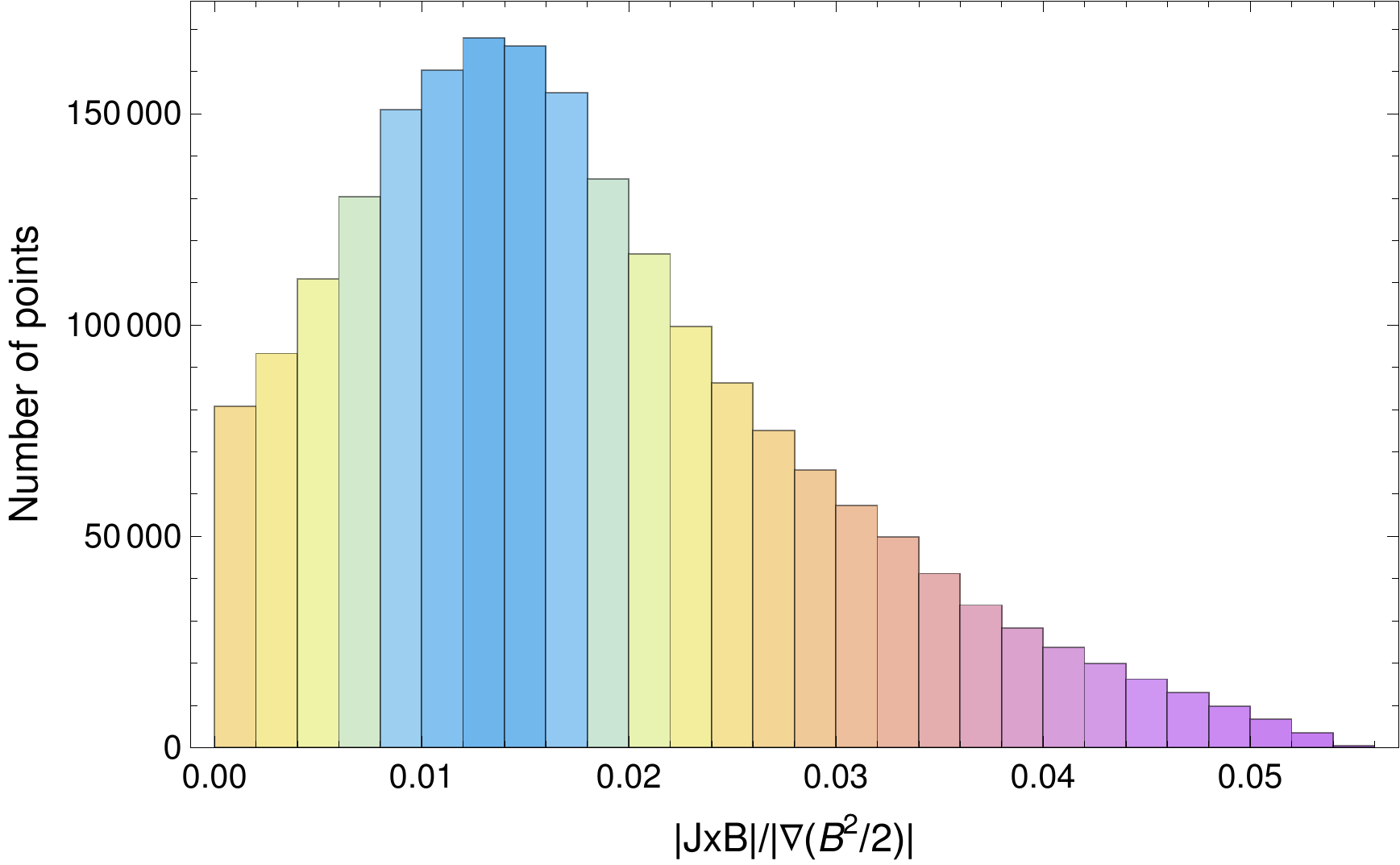}
    \caption{}
    \label{mep}
  \end{subfigure}
\quad
  \begin{subfigure}[]{0.75\textwidth}
    \centering
    \includegraphics[width=\textwidth]{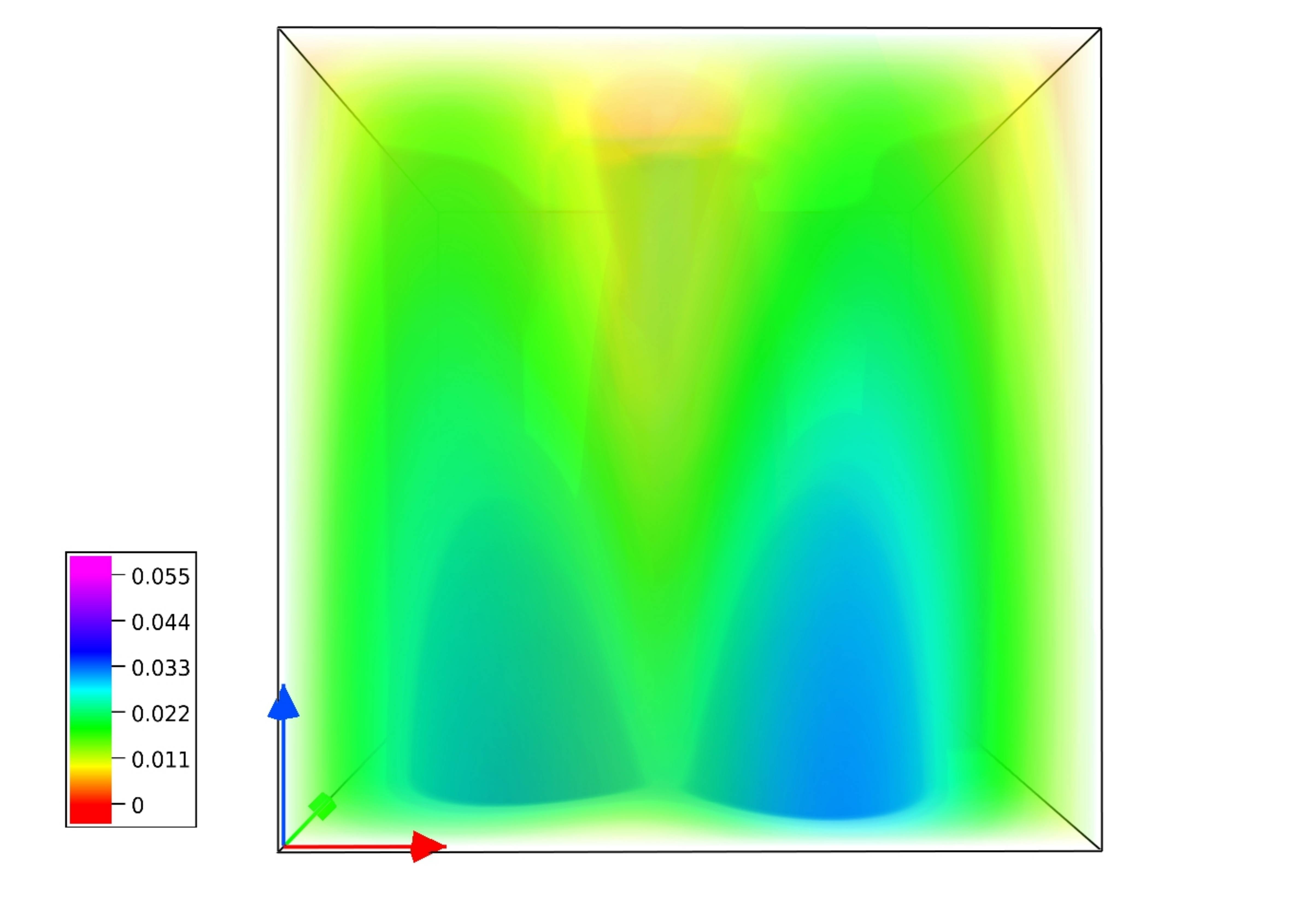}
    \caption{}
    \label{kep}
  \end{subfigure}
 \caption{Panel (a) shows the histogram of the ratio of the magnitude of the Lorentz force to the magnetic pressure gradient ($|\mathbf{J}\times \mathbf{B}|/|\nabla(B^2/2)|$) computed for the initial field. Panel (b) presents the direct volume rendering of the same over the modeling volume to show the spatial distribution of the Lorentz force. Both plots highlight the fact that the initial Lorentz force is 
small.}
 \label{f:hist}
\end{figure}

The simulations are performed on a $128\times128\times 128$ grid, resolving the normalized computational domain $\Gamma$ spanning $[0,1]\times [0,1]\times [0,0.25]$
(respectively in $x$, $y$, and $z$); see Figure \ref{f:geometry}, where a unit length is equivalent to 50 Mm.
The vertical extent of the domain is constrained by the condition that the magnetic field is to be contained within the flux surface at $r_1$ in Figure \ref{f:geometry}.
The simulations start from a motionless state with the NFFF obtained in Section 2.
 To ensure net zero magnetic flux through $\Gamma$,
the volume magnetic field $\mathbf{B}$ is continued to the domain boundaries \citep{2015PhPl...22a2902K}.
 The dimensionless constant $\tau_a / \tau_\nu \approx 7 \times 10^{-3}$, which is two orders of magnitude larger than its coronal value. To calculate ${\tau_a}$, we set $\rho_0=1$.
Notably, the larger value of $\tau_a / \tau_\nu$ only  
speeds up the computation because of a more efficient viscous dissipation by the 
increased effective drag. With time being normalized by $\tau_a$, a unit of the Alfv\'{e}n  transit time can be 
interpreted as $\approx 15$ minutes of HMI data. 
With a constant mass density, flow in the 
computation is incompressible---an assumption also used in other
works \citep{1991ApJ...383..420D,2005A&A...444..961A}. Although compressibility is important
for the thermodynamics of coronal loops \citep{2002ApJ...577..475R}, our focus is
on elucidating the changes in magnetic topology idealized with a thermally
homogeneous magnetofluid. Throughout the simulation, the maximum plasma $\beta$ varies in the 
range $\{0.003, 0.007\}$ and the fluid Reynolds number $R_F\approx 10^2$,
which reasonably approximate  the coronal values. 
The non-force-free nature of the initial field can be assessed from the histogram and the direct volume rendering of $|\mathbf{J}\times \mathbf{B}|/|\nabla(B^2/2)|$ for the initial field \citep{2012ApJ...749..135J} as shown in Figure \ref{f:hist}, where we find most of the values populated in the range $\{0.005, 0.01\}$.
Additionally, with zero resistivity $R_M=\infty$ and the flux freezing is satisfied throughout evolution, except during reconnections. 
Consequently, the MFLs evolve in response to unbalanced forces (including Lorentz force) pushing the plasma. 
The plasma motion then naturally incorporates line-tied boundaries,  which is conventional in coronal simulations \citep{schrijver2009heliophysics}. In addition, 
coronal simulations may require more sophisticated boundary treatment to replicate the observed dynamics \citep{2013ApJ...771L..30J}. 
An evolution of field lines by forcing the plasma inevitably makes the footpoints to move appreciably. A recent work concurs by suggesting  the
rapid evolution of photospheric magnetic field during major eruptive events \citep{2017arXiv170207338S}. Nevertheless, we have also performed 
simulations where the bottom boundary of the domain is kept fixed (not shown). A different flux rope, co-located with the gray colored rope (Figure 5a) 
and having similar connectivity, 
 replicates the reported evolution.

\begin{figure}[h]
    \centering
    \includegraphics[width=0.8\textwidth]{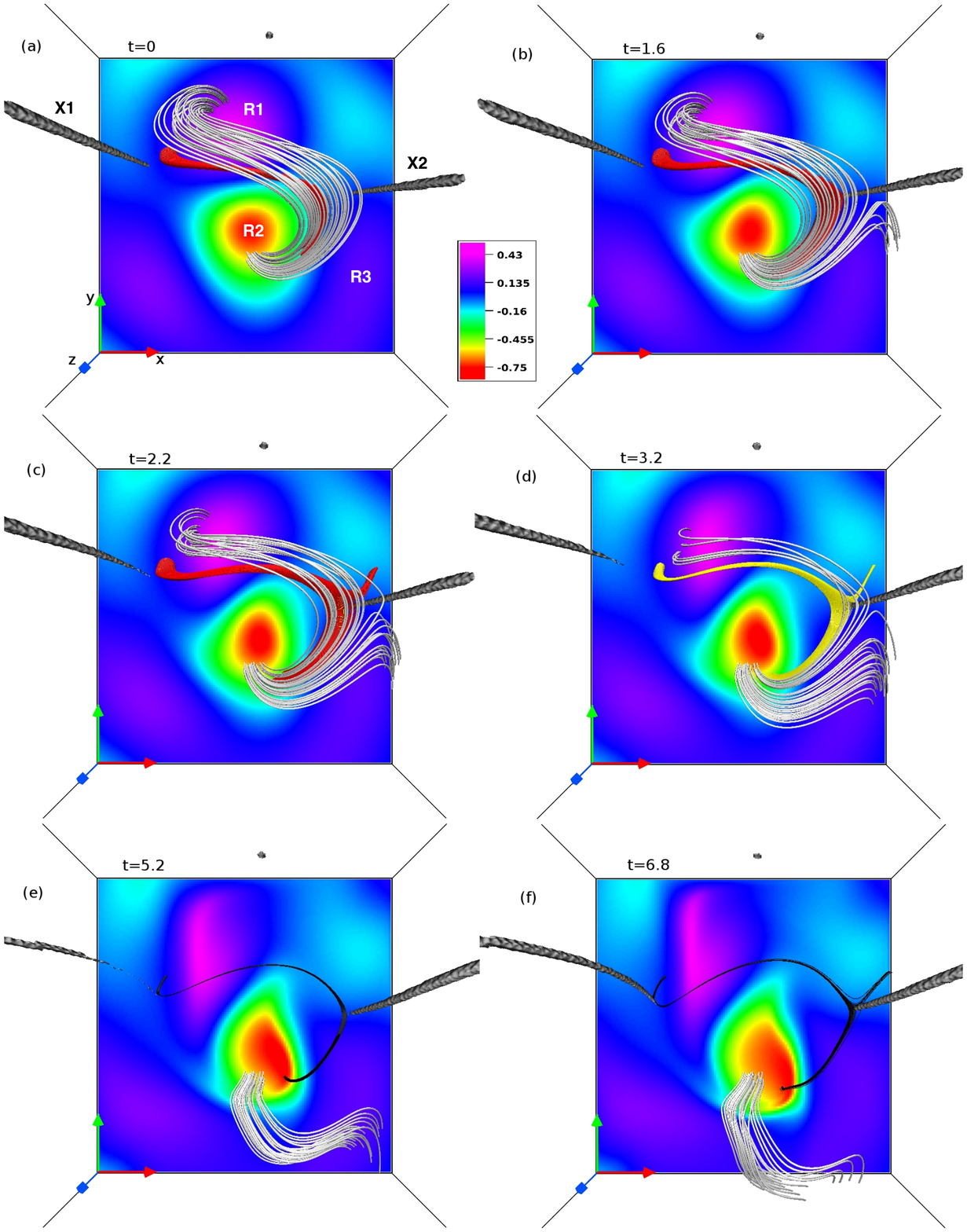}
   \caption{Evolution of the red colored rope located almost over the PIL of the major polarity regions. We also overlay the figure with overlying MFLs (in white) and magnetic nulls (in gray). The reconnections of the overlying MFLs, allowing the ascent of the structures is important. During the ascent, the local flux-surfaces constituting the structure lose their surface nature via  the reconnections of the corresponding MFLs at the null line. The main regions of field line connectivity are marked as R1, R2 and R3 in panel (a). }
 \label{f:rpt}
\end{figure}

\begin{figure}[h]
    \centering
    \includegraphics[width=0.8\textwidth]{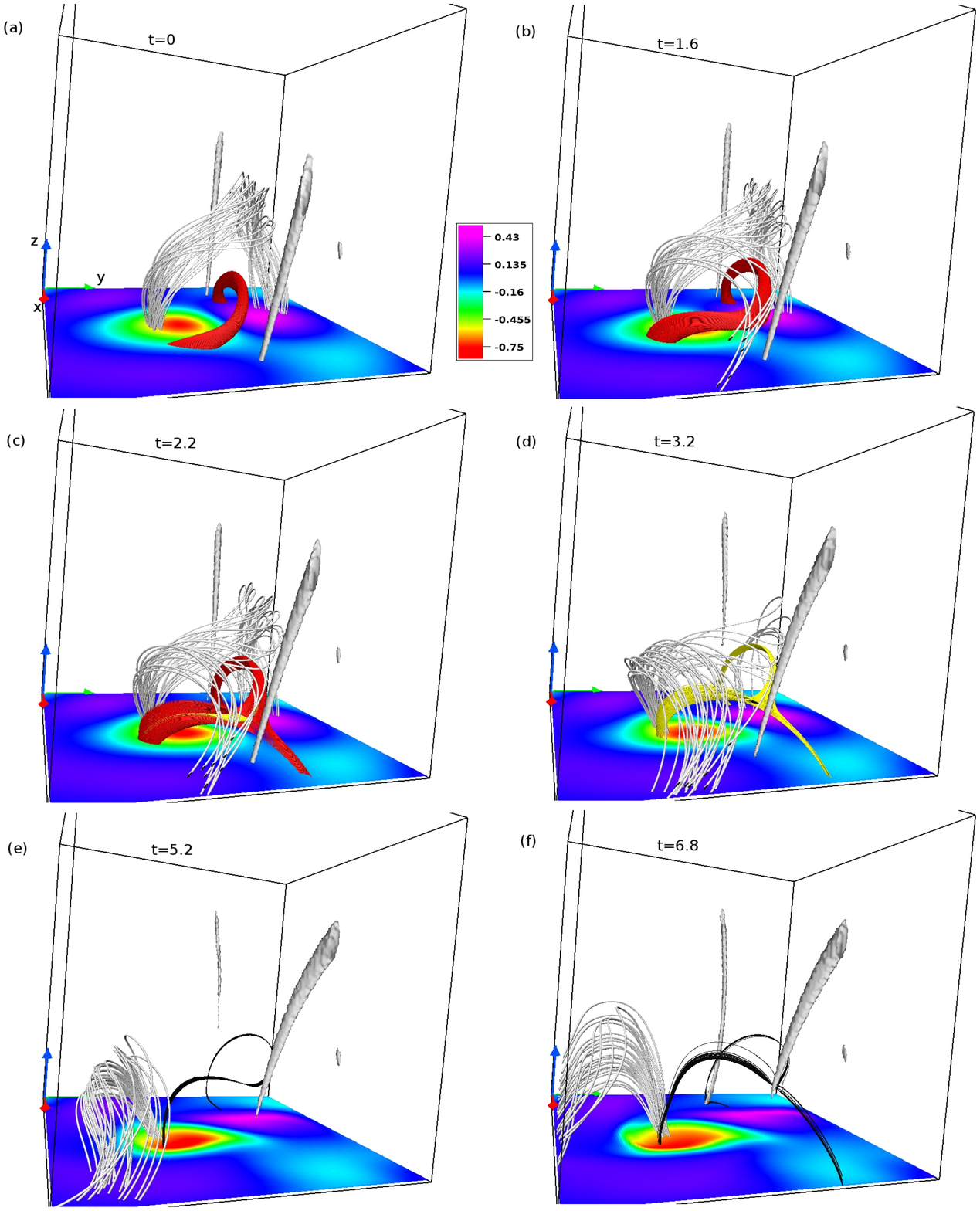}
   \caption{Side view of the rope evolution. The figure confirms the rope to be located above the major PIL 
throughout evolution.  }
 \label{f:rp}
\end{figure}

\begin{figure}[h]
  \centering
  \begin{subfigure}[]{0.45\textwidth}
    \centering
    \includegraphics[width=1\linewidth]{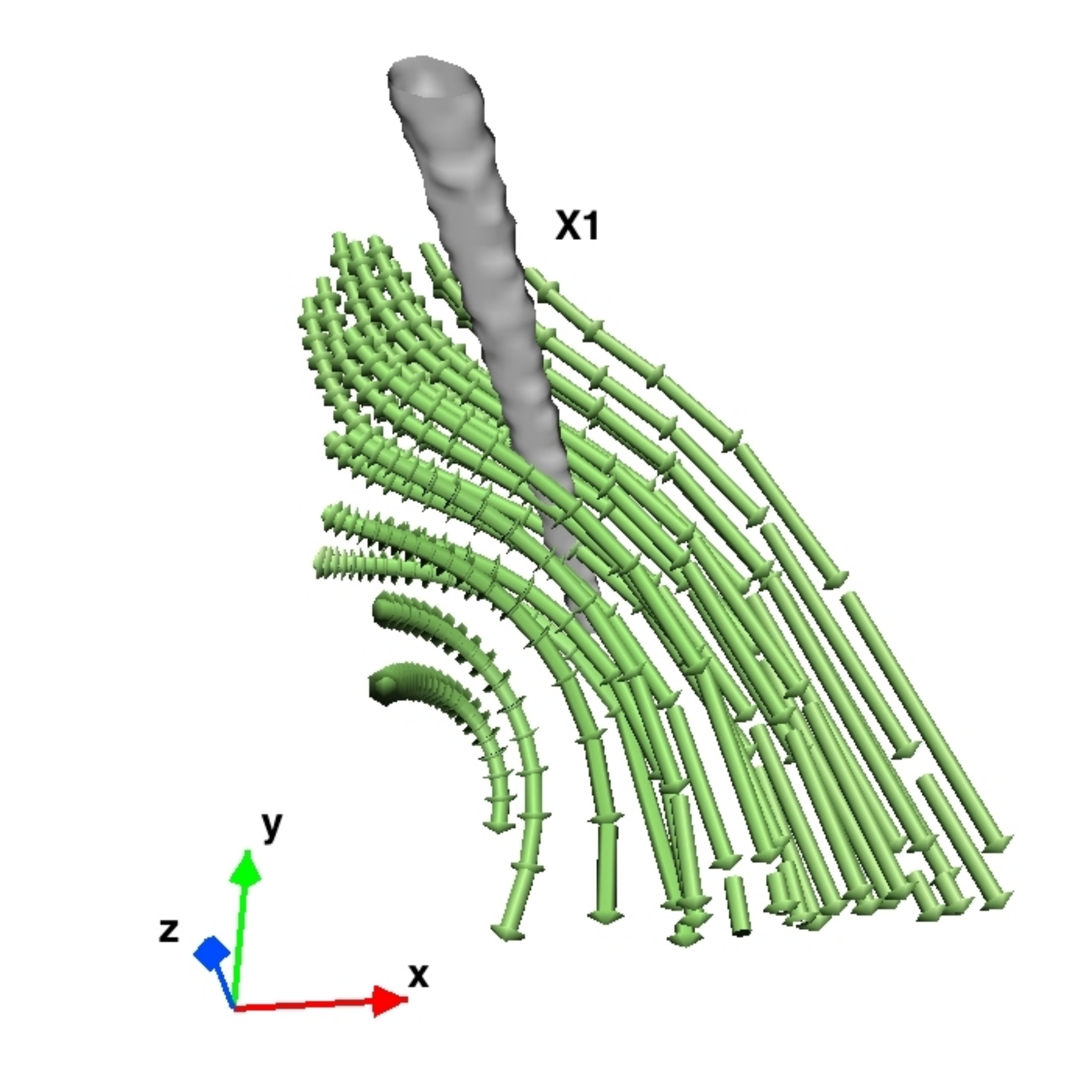}
    \caption{}
    \label{flow1a}
  \end{subfigure}
\quad
  \begin{subfigure}[]{0.45\textwidth}
    \centering
    \includegraphics[width=1\linewidth]{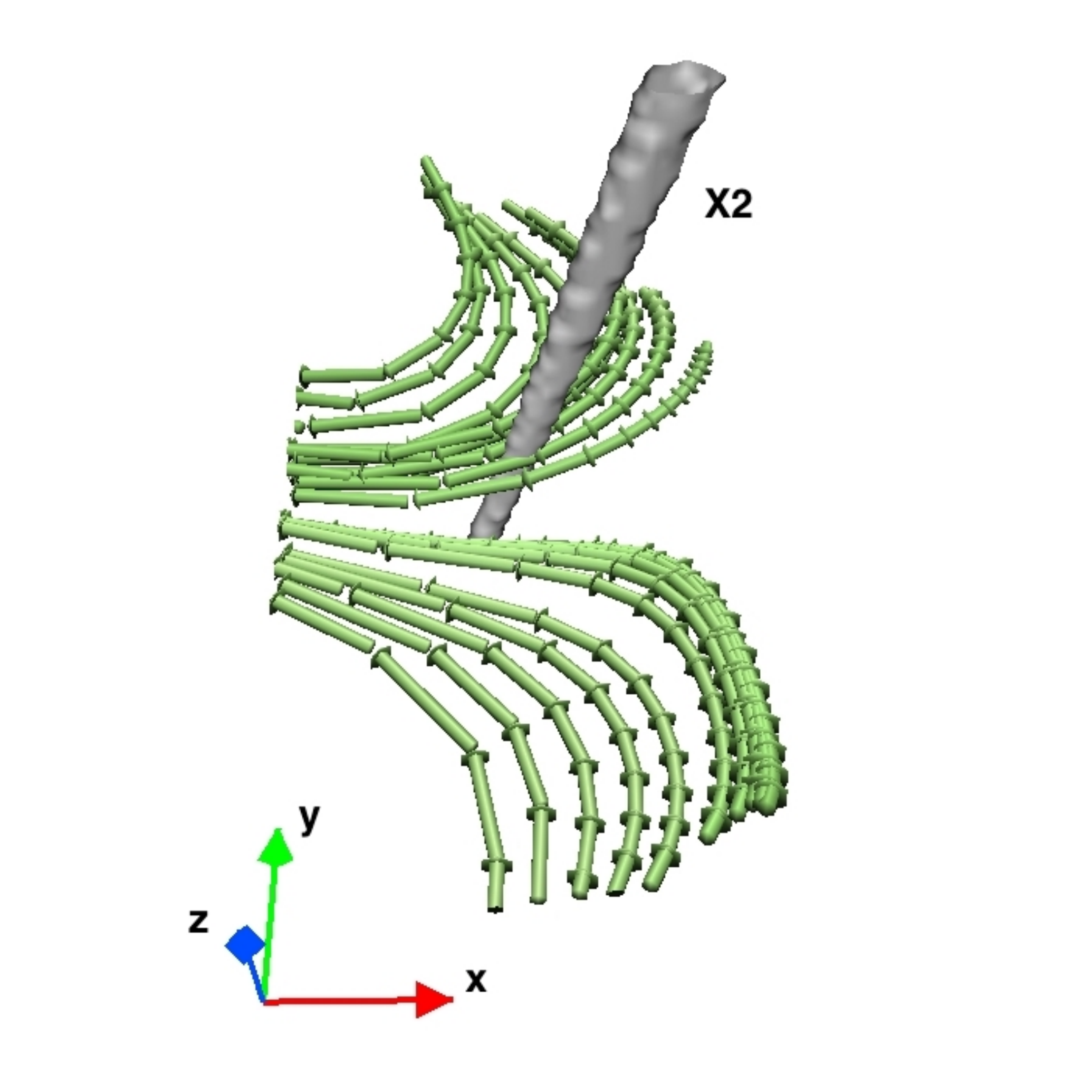}
    \caption{}
    \label{flow1b}
  \end{subfigure}
   \caption{ Panels (a) and (b) represent the streamlines of the velocity in the vicinity of X1 and X2 at $t=0.4$. The favorable orientation of the flow in the vicinity of X2, pushing to non-parallel MFLs in the close proximity to facilitate reconnection, is important. In contrast, the orientation of  the flow near X1, being less favorable, leads to a delayed onset of reconnection. 
The cumulative effect is the asymmetric ascent of the rope. }
  \label{f:flow1}
\end{figure}

Figures \ref{f:rpt} and \ref{f:rp} show the top and side views 
of a sequence of snapshots from the evolution of the overlying MFLs to the flux rope colored red in Figure \ref{f:initial_ropes}(a). Until reconnection at $t=2.2$, the rope is    
located  mostly above the main polarity inversion line (PIL) which separates the two strong opposite polarity regions R1 and R2. The figure is further overlaid with the X-type null lines X1 and X2 (in gray). Evidently, the initial MFLs of the rope are twisted while being anchored at the $z=0$ plane. 
The sequence indicates an overall anti-clockwise rotation of the overlying field lines, the rope, and the strong polarity regions. The flux rope also expands along with preferential reconnections at the null line X1. To assess the effect of rotation on the preferential reconnection,
we plot streamlines at $t=0.4$ near X1 and X2, as shown in Figure  \ref{f:flow1}.
As evident from the figure, the streamlines favor a clumping of oppositely directed field lines in the neighborhood of X2, leading to reconnections. However,  near X1, the streamlines bring field lines with the same direction toward 
the null line and hence no reconnection occurs. The net result is an asymmetric evolution that favors reconnection at X2 rather than X1. Being situated closer to the X2, the MFLs overlying the rope reconnects, first resulting in a local decrease of the magnetic tension force.

\begin{figure}[h]
    \centering
    \includegraphics[width=0.6\textwidth]{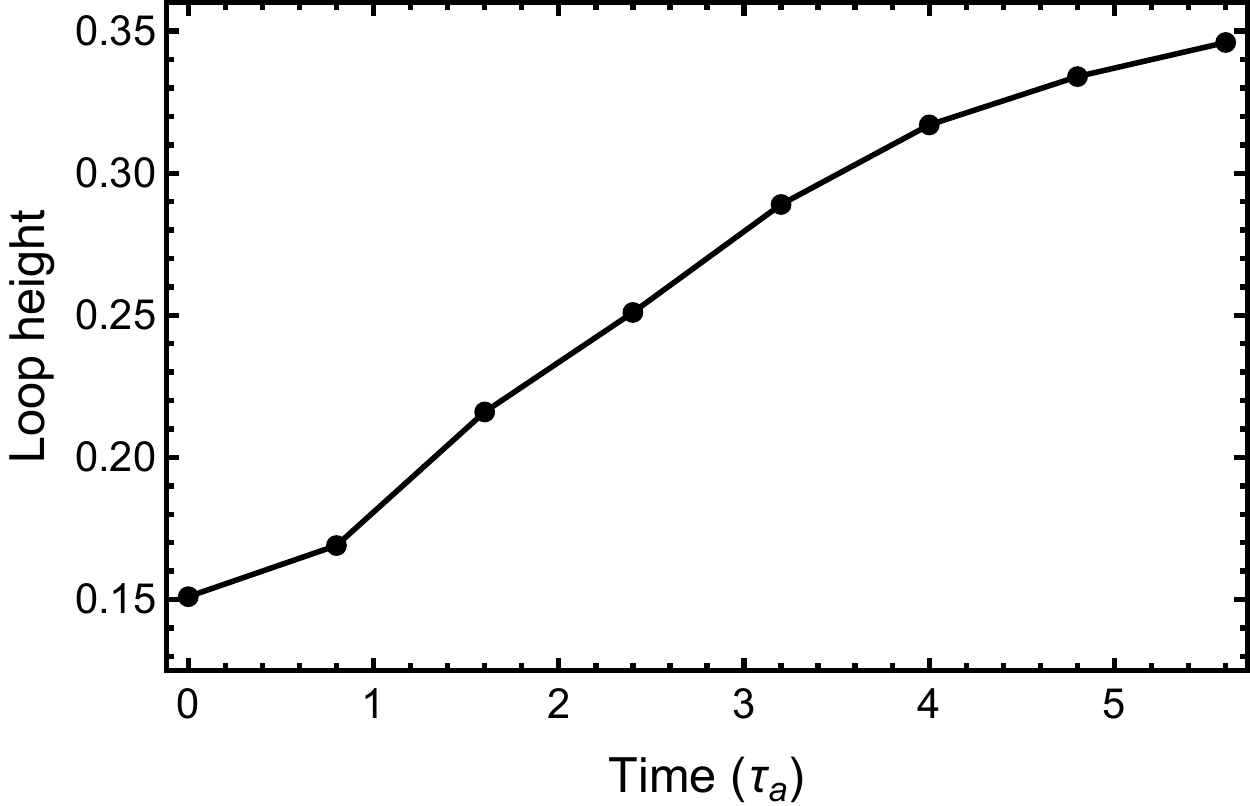}
    \caption{Monotonic increase of the height of the plotted innermost flux surface (in black, Fig. \ref{f:rp}) of the rope with time.  }
  \label{f:hp}
\end{figure}

\begin{figure}[h]
    \centering
    \includegraphics[width=0.6\textwidth]{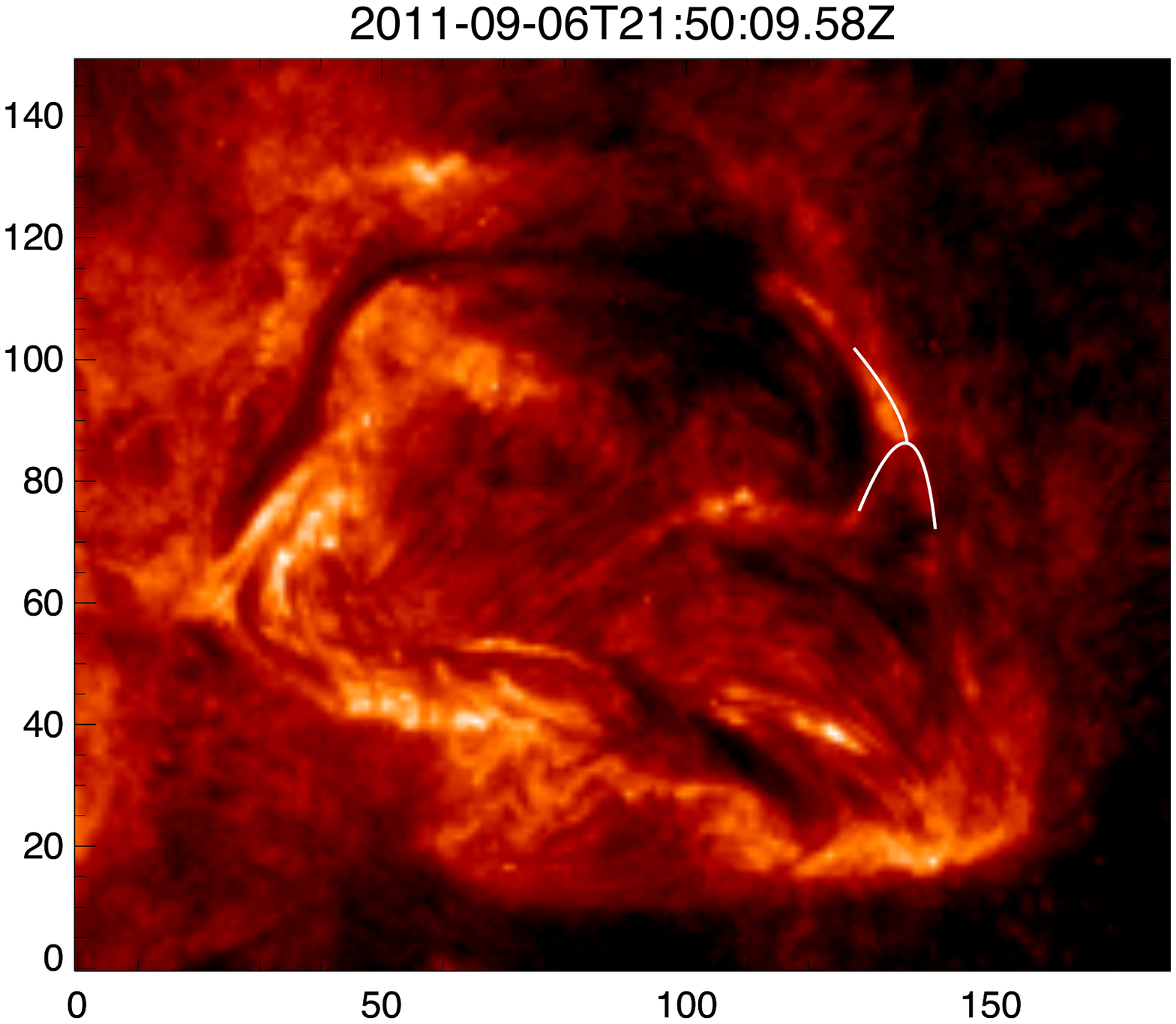}
    \caption{AIA/SDO 304\AA~ image of AR 11283 taken at 21:50 hr on 2011 September 6. The bifurcation of the filament is highlighted by white lines. The $x$ and $y$ axes represent the number of pixels in the image.
Importantly, the simulations also document flux-rope bifurcations.}
  \label{f:aia304}
\end{figure}

In response, the flux rope rises vertically while simultaneously
getting rotated toward  X2.    Figure \ref{f:hp} shows the evolution of the maximum height of the innermost flux surface of the rope (marked in black) which confirms that the rise is monotonic. Furthermore, the rise is not symmetric above the two legs of the rope 
due to the asymmetry of the initial field and the preferential reconnection. 
Notably, such a result is essentially similar to a more realistic MHD simulation initialized with an NLFFF performed by \citet{2013ApJ...771L..30J}, who show that the expansion of a low-lying flux rope is assisted by the reconnection of its overlying flux.
Furthermore, MFLs on the outermost surface of the rope (depicted in color red) reconnect, which partially opens and bifurcates the rope, making it connect region R1 with regions R2 and R3 (Figure \ref{f:rpt}). Afterwards, the inner surfaces of the rope (marked in yellow and black) also bifurcate in a similar way, leading to the  breakup of the rope. The concept of flux-rope bifurcation is well-established and responsible for the possible drainage of mass in an ascending filament \citep{2006ApJ...637L..65G,2008JGRA..113.9103G}. Importantly, the evolution of the AR 11283 also documents two-pronged pitchfork-type bifurcations of the 
associated filament when observed in the 304\AA~ channel of the AIA \citep{2012SoPh..275...17L}  at approximately UT 21:50 (Figure \ref{f:aia304}). The 
reasonably faithful simulation of the observed flux rope bifurcation justifies the credibility of the non-axisymmetric non-force-free magnetic field in exploring the coronal dynamics. However, without reproducing the bursty rise of the rope,  the present computations fail to capture the subsequent X-class flare.
The failure can be attributed  to either the exclusion of the flux emergence occurring from 22:00 hr onward \citep{2012ApJ...759...50P} in the simulated dynamics, or the requirement  of a more accurate  method to reconstruct the initial NFFF. Notably, the vertical component of the emergent field, when favorably aligned with respect to the existing field, may reconnect and provide the trigger necessary for the impulsive rise. 
 
An auxiliary analysis (not shown) indicates an increase in the magnetic flux of MFLs
overlying the ropes, in colors white and green in Figure \ref{f:initial_ropes},
because of reconnections at X2. The increased density results in an enhanced magnetic tension force which tethers the white and green ropes further and inhibit their ascent, whereas the yellow rope situated away from the main PIL shows an appreciable but non-monotonic rise because of similar tethering by reconnected field lines.

\section{Summary and conclusions}
This paper numerically simulates the viscous relaxation of a thermally homogeneous incompressible magnetofluid with infinite electrical conductivity and having an initial non-axisymmetric non-force-free magnetic field. 
The initial field is constructed by forward fitting a photospheric magnetogram using an analytically  non-force-free magnetic field. In our calculations, the focus is on  AR 11283, with magnetogram from the HMI. Notably, as the initial field is  
non-force-free and non-axisymmetric, it is able to characterize the basic structural features of such a complex AR.  The corresponding magnetic field consists of twisted field lines and two X-type null lines. Most importantly, there exists a sigmoidal flux rope that roughly matches the observed sigmoid emission. A direct visualization confirms a rope to be a stack of local MFSs. 

As the initial magnetic field is not balanced, the Lorentz force naturally trigger evolution. Subsequently, the favorable forces bring non-parallel field lines in close proximity of the null lines and lead to an unbounded sharpening of the magnetic field  gradient, which results in the development of underresolved scales. In the spirit of ILES, the MPDATA then generates locally adaptive dissipation to regularize the underresolved scales with simulated MRs.

The computation reveals, 
at one of the X-type null lines, that the initial flow favorably clumps non-parallel  MFLs and triggers MRs. The generated outflow makes the two major polarities rotate, which brings further non-parallel field lines toward the X-line and drives secondary reconnections. In response, the magnetic tension force at regions overlying the rope decreases and with the reconnections being preferential, results in an asymmetric ascent of the rope. 
Although occurring in congruence with the magnetostatic theorem, the evolution leads to reconnections occurring only at the pre-existing X-type nulls.
Importantly, the simulation depicts a sequential bifurcation of the rope over a range of time. The images from the AIA 304 \AA~ channel suggest a similar  bifurcation 
of a filament, which adds credibility to the use of non-axisymmetric NFFF to 
represent a complex AR. However, the computations fail to capture the subsequent X-class flare as the required  impulsive ascent of the rope
is missing. We attribute the failure to not accounting for the flux emergence and other solar surface modifications between the bifurcation and the flares.  For example, the recent data-driven MHD simulation  of the same AR by \citet{2016NatCo...711522J}, which includes not only the flux emergence but all other surface motions, accurately replicates the M-class flare at 01:50 hr.

In retrospect,  in its basic utilization, the non-force-free non-axisyemmteric magnetic field is useful in modeling ARs and the evolution of the solar corona. The future goal is to develop a complete numerical scheme that will simulate the dynamics of an AR from such an initial state to the occurrence of flares, if any. In the backdrop of the successful simulation of coronal dynamics by \citet{2016NatCo...711522J}, we believe that a synergetic approach exploiting the advantages of various numerical models is the key to achieve this goal. For instance, the fully numerical non-force-free extrapolation model developed by 
\citet{2008SoPh..247...87H} and its subsequent developments \citep{2008ApJ...679..848H,2009SoPh..257..271G}, rely on self-organized higher-curl Beltrami fields \citep{2007SoPh..240...63B,2011PhPl...18h4506K,2012PhPl...19c2113J} that are
 morphologically identical to the initial magnetic field $\mathbf{B}$. 
Therefore, it can be expected that a data-driven MHD simulation of reconnections with the initial field provided by non-force-free extrapolations will provide a more realistic dynamics of the coronal magnetic field and thus warrants further attention.

\acknowledgements 

\noindent {\bf Acknowledgements:} The simulations are performed using the 100 TF
cluster Vikram-100 at the Physical Research Laboratory, India.
We acknowledge the use of the visualization software VAPOR (www.vapor.ucar.edu) for generating relevant graphics. Data and images are courtesy of NASA/SDO
and the HMI and AIA science teams. SDO/HMI is a joint effort of
many teams and individuals to whom we are greatly indebted for
providing the data. The authors are thankful to Dr. P. K. Smolarkiewicz for his invaluable contribution toward the better presentation of this paper. We are also thankful to the anonymous referee for providing insightful suggestions toward the overall betterment of this paper.

\bibliography{ms}

\newpage
\newpage
\newpage

\newpage

\newpage
\newpage

\newpage

\newpage
\newpage

 \newpage

\newpage

\newpage

\end{document}